\documentclass[prl,nofootinbib,twocolumn,showpacs]{revtex4}
\usepackage{graphicx}%
\usepackage{dcolumn}
\date{today}
\def\bq{\begin{eqnarray}}
\def\ee{\end{eqnarray}}
\begin{document}
\begin{abstract}
\noindent We analyze some consequences of the Casimir-type zero-point radiation pressure.
These include macroscopic "vacuum" forces on a metallic layer in-between a dielectric
medium and an inert ($\epsilon (\omega) = 1$) one. Ways to control the sign of these
forces, based on dielectric properties of the media, are thus suggested. Finally, the
large positive Casimir pressure, due to surface plasmons on thin metallic layers, is
evaluated and discussed.
\end{abstract}
\title{The Casimir zero-point radiation pressure}
\author{Yoseph Imry} \affiliation{Dept. of Condensed-Matter
Physics, the Weizmann Institute of Science, Rehovot 76100,
Israel} 
\date{\today}
\maketitle

Imagine polarizable bodies that are placed in vacuum. Their interaction
with the electromagnetic field (which  can often be described by boundary conditions on
the latter at the surfaces of the bodies) may produce a change in the
zero-point energy of the field.  Should that energy  depend, for example, on the distance
between two of these bodies,  forces between these two bodies will follow. This can
be regarded as the origin of the van der Waals molecular forces
\cite{London}, which change at large separations due to
retardation effects \cite{Casimir-Polder}. For the simpler case of two large parallel
conducting plates, the Casimir force \cite{Casimir} (cf. Eq. \ref{Cas} below) results at
large
separations (where retardation is important) between the plates , and becomes the
Lifshitz  force \cite{Lifshitz,Dzy} at small separations (where
quasistationarity applies). The crossover between the short- and
long-distance behaviors occurs for distances  on the order of the
velocity of light divided by the characteristic excitation
frequency of the bodies ({\ i.e.} about $200 A$ for $\hbar \omega = 10 eV$). Even for a single
body, volume- and shape \cite{B-dP}-dependent forces will arise when the field energy
depends on these parameters. The Casimir force has by now been
amply confirmed by experiment \cite{exp}. Corrections due to finite
temperatures, realistic surfaces, etc. are becoming relevant
\cite{Genet}. The Casimir effect may be crucial to nanomechanical
devices \cite{mech}. Its relevance is not limited to the electromagnetic
field only. It should exist with any physical bosonic field that interacts with
matter.

Besides its general interest vis a vis the observability of {\em
(only changes of)} the vacuum energy \cite{rem} and genuine relevance to
molecular and colloidal forces, the Casimir effect touches upon
several fundamental questions of Physics. These range from "vacuum
friction" to the value of the cosmological constant and the modifications of
classical Newtonian gravitation on small scales. The reader is
referred to several books and review articles,
which discuss the many aspects of the Casimir effect
\cite{Milonni-book,Most-book,Bord-Phys-Rep,B-G,P-M-G,Milton}.

A problem of principle which arises in the calculation of Casimir-type forces is the
well-known UV divergence of the electromagnetic vacuum energy. This divergence is clearly
physically irrelevant here, since what matters are only {\em differences} of energies.
For a good discussion of the cutoff procedure see \cite{cardy}. Ordinary metals are
basically transparent at high frequencies,  above the characteristic plasma frequency
$\omega_p$ which is therefore a natural cutoff. It is clear that waves with $\omega \gg
\omega_p$ do not "see"  the bodies and therefore are irrelevant. In his original
calculation Casimir in fact first employed a soft cutoff as above and then made a
judicious subtraction of a large energy to obtain a finite, universal and
cutoff-independent result. This subtraction procedure is rather tricky. Although various
physical interpretations for it have been suggested in the literature (see below), none
of them is truly satisfactory.  We shall start by physically analyzing Casimir's
subtraction procedure. Before that, we remark that cutoff-dependence can be allowed when
the cutoff is based on physical considerations. For example, the Lifshitz forces in the
static limit do depend on the cutoff $\omega_p$, where $\omega_p$ is the plasma frequency
of the metals. Another example of cutoff-dependence will be discussed in this paper.

In 1948, Casimir \cite{Casimir} considered the force between two large metallic plates
placed parallel to the x-y plane, with a distance $d$ along the $z$ axis between their
internal faces, and $d \gg c/ \omega_p$. The zero-point energy of the field between the
plates is

\bq E_0(d) = \hbar c \frac{L^2}{\pi^2}\int^{(c)} d^2 k_\bot \sum_{(0)} ^\infty (n^2
\frac{\pi^2} {d^2} +  k_\bot ^2)^{1/2}, \label{E} \ee

\noindent where $\int^{(c)}$ means that the integrand is multiplied by a soft
cutoff-function which vanishes smoothly around and above $|k_p| = \omega_p / c$, and $\sum_{(0)}
^\infty$ means that the $n = 0$ term is multiplied by $1/2$. The corresponding subtracted
quantity is:

\bq E'_0(d)  = E_0(d) -  subtraction. \label{E'} \ee

\noindent The  force between the plates is given by

\bq F = - \frac{\partial E'_0(d)} {\partial d}, \label{F} \ee where positive F means
repulsion between the plates. Casimir chose to subtract in Eq. \ref{E'}  the same
expression  but with the sum over $n$ converted to an integral, as appropriate for very
large $d$. Thus, the subtraction is that of the energy for the plates "at infinity"
(questions such as whether the plates have a finite thickness, and if so -- what happens
beyond them, are left open). Evaluating the difference between the sum and the integral
over $n$ with the Euler-Maclaurin formula, he arrived at the celebrated result:

\bq
P_c = F/L^2 = -\hbar c \frac{\pi^2}{240} \frac{1}{d^4}.\label{Cas}
\ee

\noindent For  the unretarded, quasistationary limit, $d \ll c/\omega_p$, a length $\sim
c/\omega_p$ replaces one power of $d$ in the denominator of Eq. \ref{Cas}, as found by
Lifshitz \cite{Lifshitz}.

A clear physical justification for the subtraction procedure is clearly called for. It is
immediately suggested \cite{Milonni-pressure} (and in fact hinted in Casimir's original
paper) that the physical significance of the above subtraction is in obtaining the
difference between the radiation pressures of the zero-point fields between the plates,
and  outside of the plates. This idea was advocated and followed up in Ref.
\cite{Milonni-pressure}. The purpose of this paper is to analyze some new consequences of
this interpretation of the subtraction. Neither it nor the other regularization
procedures are truly satisfactory. Therefore, it is of interest to compare the new
results following from this interpretation of the subtraction procedure with experiments
to come.

We follow  Ref. \cite{Milonni-pressure} in calculating the pressure of the zero-point EM
field, but present here a somewhat diferent derivation. We take a large vessel
\cite{vessel} with reflecting walls. The vessel is taken  to be a box with dimensions
$L_x, L_y, L_z$, $V = L_x L_y L_z$. The pressure in the z-direction is given by the
momentum imparted to the wall per unit area per unit time \cite{Milonni-pressure}:

\bq P_0 = \hbar \sum_{k_x,k_y,k_z}^{(c)} c(k) \frac{k_z^2}{k} /V, \label{pressure}\ee

\noindent where $k_x = n_x \pi /L_x$, etc. and c(k) is the light velocity as a function
of $k \equiv \sqrt{k_x^2 + k_x^2 + k_z^2}$, slightly generalizing the result of Ref.
\cite{Milonni-pressure}. The symbol  $(c)$ above the summation sign signifies an upper
cutoff around the plasma frequency of the walls, necessary to control the divergence, as
discussed above. To derive this result, one may calculate \cite{check} $-\frac{\hbar
}{L_y L_z}\frac{\partial (c k)}{\partial L_z} = \frac{\hbar c k_z^2}{k V}$ and sum over
the levels. The possible volume dependence of $c(k)$ is neglected.

For a large system, the sum can be replaced by an integral, we perform the angular
integrations and change variables from $k$ to frequency ($\omega$), obtaining:

\bq P_0 = \frac{\hbar}{6 \pi^2 c^3} \int^{(c)} d \omega  \omega^3 \epsilon(\omega)^{3/2}, \label{pressure-omega}\ee

\noindent where the factor of $1/2$ in the zero point energy and the degeneracy of each
$k$ mode cancelled and we used the frequency-dependent $\epsilon(\omega)$ via $c(\omega)
= c / \epsilon(\omega)^{1/2}$. The superscript $(c)$ signifies an upper cutoff around the
plasma frequency of the walls, as above.
 Defining $\overline{c}$ as a
suitable average of $c(\omega)$, (i.e. $\frac{1}{\overline{c}^3} = \frac{ \int^{(c)} d
\omega \omega^3 \epsilon(\omega)^{3/2}}{\int^{(c)} d \omega \omega^3 }$), one obtains

\bq P_0 \cong  \frac{\hbar \omega_p^4}{24\pi^2 \overline{c}^3}.\label{press}
\ee

We used an approximate equality, since we gave the result  for a sharp cutoff and a soft
cutoff may change it somewhat.

One might \cite{confess} try to use here (as in Eq. \ref{F}, based on \cite{Casimir}) the   thermodynamic relationship,   for a system which
in equilibrium at $T = 0$ (see also Ref. \onlinecite{Bart}):

\bq P_0 = - \frac{\partial E_0} {\partial V} = - \frac{\partial} {\partial V} \int^{(c)} V d \omega \frac{\omega^2}{\pi^2
c^3} \frac{\hbar
\omega} {2}. \label{therm}\ee

\noindent This would produce a
 {\em negative} pressure. However \cite{confess}, this relationship is valid only for a
 closed system. But  the present system exchanges energy with the continuum levels above
 $\omega_p$,
 when its volume varies, via zero-point photon levels moving through the cutoff.
 Interestingly, Casimir  used the same relationship to calculate the net pressure on
 each plate. We believe that this may be justified for pressure differences, but
 {\em only} when the media on the two
 sides of each metallic plate are equivalent. This point will be more fully discussed elsewhere.

The pressure of Eq. \ref{press} is not-so-large but quite significant. It is convenient
to express it in terms of a Bohr (or Fermi) pressure $P_B \simeq  10eV / A^3 \simeq 1.5
\cdot 10^8 N/cm^2$. For $\hbar \omega_P = 10eV$, we find $P_0 \sim 1.5 \cdot 10^{-9} P_B
\sim 0.2 N/cm^2$. For comparison, the ordinary Casimir force/unit area at a distance of
$100 nm$ is on the order of $10^{-3}N/cm^2$.

Since the  ordinary Casimir force is the result of the  near-cancellation of much larger
quantities, its sign is notoriously difficult to predict, except via detailed
calculations \cite{Boyer}. We suggest that some control of the sign can be achieved \cite{remark} by
employing polarizable materials as the electromagnetic vacuum in some part of the system.
For a material with a dielectric constant $\epsilon(\omega)$, Eq. \ref{pressure-omega}
suggests that if the suitably averaged value, $\overline{\epsilon} > 1$ (where
$\overline{c} \equiv c/\sqrt{\overline{\epsilon}}$), which should often happen, the
pressure of the dielectric will exceed that of the vacuum by $\Delta P $:

\bq \Delta P \cong  \frac{\hbar \omega_p^4}{24\pi^2} (\frac{1}{\overline{c}^3} - \frac{1}{c^3}).
\label{diff}\ee

Thus, for example, a metallic wall having a dielectric medium with such an
$\epsilon(\omega)$ on one side and a medium with $\epsilon = 1$ on the other, both having
the same mechanical pressure,  will be attracted into the vacuum \cite{el}. As a weak
example we take $\epsilon(\omega) =2$ up to $0.05$ of the metallic $\omega_p$, this will
give a net force per unit area of $~10^{-6} N/cm^2$. In addition, one may think about
macroscopic vessels either filled with or immersed in a dielectric fluid with the same
mechanical pressure as the inert, $\epsilon = 1$, material outside/inside. It appears
possible to observe, in principle with an interferometric method,  the small changes of
their macroscopic dimensions between these two situations, for example. This would
constitute a {\em macroscopic} version of the Casimir effect.

Things become rather interesting also for the ordinary, mesoscopic-scale, Casimir effect.
A good check of the present interpretation of the Casimir subtraction is the following:

Consider the case where the medium outside the plates is "inert" ($\epsilon = 1$) and the
medium between them has an $\epsilon(\omega)$, with $\overline{\epsilon} > 1$. The
conventional calculations treat the case in which these two media are identical (with the
same $\epsilon(\omega)$) . Let us then start with both the inside and outside media
identical and having an $\epsilon(\omega)$. The Casimir pressure in this case was
calculated in Ref. \cite{Dzy}. We denote it by $P_c(\epsilon)$. In the case of interest
to us the medium outside is inert, so we have to subtract the pressure of the vacuun
rather than the pressure of the dielectric medium. We then find that the net Casimir
pressure is, in our case:

\bq P_c (1~inside, \epsilon~outside)= P_c(\epsilon) - P_0(\epsilon = 1)  + P_0(\epsilon).
\ee

\noindent Therefore, for sufficiently large $\epsilon(\omega)$ the sign of the force will change
and it will  push the plates away! For the aforementioned example, considered below Eq.
\ref{diff}, this repulsion will win against the Casimir attraction around a distance of
about $0.6 \mu$. {\em For larger distances, the full Casimir force should ideally be repulsive}; see however \cite {el}.
This change of sign is due to the larger "volume force" due to the dielectric inside.

At distances below $c/\omega_p$, where quasistationarity holds, the outside pressure
$P_0$ may again be smaller than the inside Lifshitz pressure. Interesting effects due to
dielectric media placed between or outside of the plates are possible and will be
discussed elsewhere.

We conclude this note by examining the Casimir vacuum forces on a single flat metallic
plate of thickness $d$. For large thicknesses, we simply have the two  pressures, $P_0$,
from  the two sides of the metallic layer. These will slightly decrease the thickness of
the layer, a very interesting effect which can be increased with dielectric materials as
discussed above and might be observable some day. In addition to the ordinary
electromagnetic modes considered so far, there will be surface plasmons
\cite{Stern-Ferrel,Raether},\cite{Lifshitz,Dzy,van Kampen} running on the two interfaces
of the layer. For a thick layer, the energy of these modes  will be independent of $d$,
but once $d$ becomes comparable to the decay-lengths of the modes, their energies will
depend on $d$ and lead to a significant further positive pressure on the metallic plate.

To calculate that pressure, we consider a metallic slab with a dielectric constant
$\epsilon(\omega) = 1 - \frac{\omega_p^2}{\omega^2}$ and  of thickness $d = 2a$, larger
than atomic dimensions, between
the planes $z = \pm a$. Following Ref. \cite{van Kampen}, we approximate  in the
quasistationary limit the full wave equation by the Laplace one for the electrostatic
potential $\phi$. We take without loss of generality a wave propagating in the x
direction, $\phi(x,z) =  exp(ikx)u(z)$, and find $u'' = k u$. Thus $u = \sum_\pm A_{\pm}
exp( \pm kz)$ inside the film and u is exponentially decaying in the two vacua (with
$\epsilon = 1$) on the two sides of the film. On the surfaces of the film $\phi$ and
$\epsilon \frac{\partial \phi}{\partial z}$ are continuous. By symmetry, we choose even
and odd solutions with respect to $z = 0$, and find the surface plasmons' dispersion
relations: \bq \omega_{\pm}(k) = \frac{\omega_p}{\sqrt{2}}\sqrt{1 \mp e^{-kd}}, \ee

\noindent where the upper/lower sign is for the even/odd modes. In the extreme
quasistationary limit, $d \ll c/\omega_p$, we may neglect the polariton effect -- the
coupling of the above modes with the "light modes" $\omega = ck$. The dispersion of the
latter is extremely steep and intersects the $\omega_{\pm}(k)$ dispersion only at very
small values of $k$.

To obtain the force one needs the derivative with respect to $d$ of the $d$-dependent
total zero-point energy of these plasmons. One may either directly take the derivative
with respect to $d$ or first integrate the energies subtracting from each branch an
infinite d- independent constant, which is the $k \rightarrow \infty$ limit of both
dispersion curves:
 \bq
 E_0(d) = \frac{1}{2} \sum_\pm \hbar (\frac{L}{2\pi})^2\int d^2 k (\omega_{\pm}(k) - \frac{\omega_p}
 {\sqrt{2}}),\\
F(d) = - \frac{\partial}{\partial d} E_0(d).
 \ee

\noindent In both ways, we find for the pressure (which turns out to be positive) exerted
by the vacua on the metallic film, a result resembling the Lifshitz pressure in the non-
retarded regime \cite{Lifshitz,Dzy}: \bq P(d) = \frac{F(d)}{L^2} = 0.0078 \frac{\hbar \omega_p}{d^3}
\label{film}. \ee  This pressure is quite substantial and increases markedly with
decreasing d. It is on the order of $2\cdot10^6 N/cm^2$ for a $~1 A$ thin film -- almost
approaching the Fermi pressure scale for atomic thicknesses. The Fermi (including the
Coulomb) pressure will eventually stabilize the very thin layer  against squeezing by the
vacuum pressure. These considerations  are clearly relevant for the Physics of very thin
films. More work is needed to check their relevance  elsewhere.

To summarize, we considered the radiation pressure of bulk zero-point EM modes. The
dependence of the force on the dielectric constant of the electromagnetic vacuum leads to
a novel type of force in asymmetric situations where the  conducting slab has different
dielectrics on its two sides.  Options for controlling the sign in the Casimir-type
geometry are suggested. Finally, the substantial positive pressure, associated with the
surface plasmons, exerted by the electromagnetic vacuum on on a thin metallic film  was
evaluated and discussed.

\vspace{0.5cm}
\noindent{\bf Acknowledgements}\\

\noindent The author thanks  M. Aizenman, A. Aharony, T. Emig, O. Entin-Wohlman, U. Gavish, Y. Levinson,  S. Rubin, A. Schwimmer, A. Stern and Z. Vager,
and especially W. Kohn and M. Milgrom, for discussions and comments on the manuscript. This work was supported by  a Center of
Excellence of the Israel Science Foundation and by the German Federal Ministry of
Education and Research (BMBF), within the framework of the German-Israeli Project
Cooperation (DIP).


\begin{thebibliography}{99}
\bibitem{vessel} On the right-hand side (RHS) of the Casimir geometry this vessel is usually taken to consist of the RHS infinite plate and another parallel infinite plate very far away on the right.
\bibitem{check} For the $n$'th photon eigenstate of an infinite  square well of size $L$ along
the $z$ axis, a momentum $\frac{2 \hbar n \pi}{L}$ is given to each wall in each period
$2L/c$. This is identical, as it should, to (minus) the derivative with respect to $L$ of
the energy
$\frac{ \hbar cn \pi}{L}$.
\bibitem{remark} A change of sign, due to different signs of dielectric function differences with the middle layer, has been found in ref. \cite{Dzy} in the asymmetric three-layer situation. It appears to be due to a different physical origin from the change of sign discussed here in the symmetric five-layer situation.
\bibitem{el} This force will lead to a finite displacement of the slab,  balanced by
the elasticity of the medium. See also \cite{Lifshitz,Dzy,Lev}. It may be possible to observe the displacement of the slab's position and  interfaces by changing the dielectrics.
\end{thebibliography}
\end{document}